\documentstyle[12pt]{article}

\def\n{\hfill\break}

\def\tenbf{\normalsize\bf}
\def\tenrm{\normalsize\rm}
\def\tenit{\normalsize\it}
\def\elevenbf{\normalsize\bf}
\def\elevenrm{\normalsize\rm}

\def\vereq#1#2{\lower3pt\vbox{\baselineskip1.5pt \lineskip1.5pt
\ialign{$\m@th#1\hfill##\hfil$\crcr#2\crcr\sim\crcr}}}

\def\to{\rightarrow}

\def\eslt{E\llap/_T}
\def\Re{{\cal R \mskip-4mu \lower.1ex \hbox{\it e}}\,}
\def\Im{{\cal I \mskip-5mu \lower.1ex \hbox{\it m}}\,}

\def\tg{\tilde g}

\def\tnu{\tilde\nu}

\def\tq{\tilde q}
\def\tst{\tilde t}

\def\tw{\widetilde W}
\def\tz{\widetilde Z}
\def\tf{\tilde f}
\def\tl{\tilde l}
\renewenvironment{thebibliography}[1]
 { \elevenrm
   \begin{list}{\arabic{enumi}.}
    {\usecounter{enumi} \setlength{\parsep}{0pt}
     \setlength{\itemsep}{3pt} \settowidth{\labelwidth}{#1.}
     \sloppy
    }}{\end{list}}

\begin{document}
\begin{flushright}
FSU-HEP-930329 \\
UH-511-764-93 \\
April 1993
\end{flushright}
\begin{center}{{\tenbf SIMULATING SUPERSYMMETRY WITH \\ ISAJET 7.0 / ISASUSY
1.0}
\vglue 5pt
\vglue 1.0cm
{\tenrm HOWARD BAER \\}
\baselineskip=13pt
{\tenit Department of Physics, Florida State University\\}
\baselineskip=12pt
{\tenit Tallahassee, FL 32306 USA\\}
\vglue 0.5cm
{\tenrm FRANK E. PAIGE \\}
\baselineskip=13pt
{\tenit Superconducting Supercollider Laboratory\\}
\baselineskip=12pt
{\tenit Dallas, TX 75237 USA\\}
\vglue 0.5cm
{\tenrm SERBAN D. PROTOPOPESCU\\}
\baselineskip=13pt
{\tenit Brookhaven National Laboratory\\}
\baselineskip=12pt
{\tenit Upton, NY 11973 USA\\}
and
\vglue 0.5cm
{\tenrm XERXES TATA \\}
\baselineskip=13pt
{\tenit Department of Physics and Astronomy, University of Hawaii\\}
\baselineskip=12pt
{\tenit Honolulu, HI, 96822 USA\\}
\vglue 0.8cm
{\tenrm ABSTRACT}}
\end{center}
\vglue 0.3cm
{\rightskip=3pc
 \leftskip=3pc
 \tenrm\baselineskip=12pt
 \noindent
We review the physics assumptions and input used in ISAJET~7.0 /
ISA\-SUSY~1.0 that are relevant for simulating fundamental processes
within the framework of the Minimal Supersymmetric Standard Model (MSSM) at
$p\bar p$ and $pp$ colliders.  After a brief discussion of the
underlying MSSM framework, we discuss event simulation and list the
sparticle production processes and decay modes that have been
incorporated into our calculations. We then describe how to set up and
run an ISAJET / ISASUSY job and the user input and output
formats. The ISAJET program is sufficiently flexible that some
non-minimal supersymmetry scenarios may be simulated as well.
Finally, plans for future upgrades which include the extension to $e^+
e^-$ collisions, are listed.
\vglue 0.6cm}
{\elevenbf\noindent 1. Introduction}
\vglue 0.2cm
\baselineskip=14pt
\elevenrm
There are many reasons for believing that supersymmetry --- a symmetry
between fermionic and bosonic degrees of freedom --- may be an actual
symmetry of nature\cite{MSSM}. One reason, in particular, is that in
the Standard Model (SM), the instability of elementary scalar fields
to radiative corrections leads to the well known fine-tuning problem.
Supersymmetry (SUSY) provides the only known framework that allows for
the introduction of elementary scalar fields, essential for the
breaking of electroweak symmetry, into the theory, without the need
for fine tuning parameters to uncanny accuracy. This, however,
requires that the SUSY breaking scale is ${\cal O}$(1 TeV), so that
the supersymmetric partners of known particles should be accessible at
high energy hadron colliders.

How does one make contact between the theoretical ideas of
supersymmetry and gauge theories of quarks and leptons on the one
hand, and the observation of real particles such as $\pi$'s, $K$'s,
$e$'s, $\mu$'s and $\gamma$'s in complicated collider experiments on
the other? A crucial bridge of this gap has been provided by the
development of event generation and simulation programs\cite{SJO} such
as ISAJET\cite{ISA}.  Such programs merge perturbatively calculable
hard scattering processes with approximate all-orders QCD corrections
and non-perturbative models for the hadronization of quarks, gluons
and beam fragments. Programs have been developed to simulate
essentially all lowest order and some higher order SM processes along
with a few processes arising from physics beyond the SM. The capacity
to simulate production of supersymmetric particles, however, has been
included only at an elementary level in some of the generators
reviewed in Ref.~\cite{SJO}. Motivated by both theoretical as well as
aesthetic considerations, we have made a concerted effort to include a
more realistic simulation of supersymmetry in ISAJET version 7.0, so
that the experimental consequences of supersymmetry can be explicitly
viewed in the environment of a collider detector.

In this report, we describe the theoretical structure of supersymmetry
incorporated into ISAJET 7.0, and we explain how to set up
and run the program, so that users may simulate production and decays
of supersymmetric particles of the Minimal Supersymmetric
Model\cite{MSSM} at hadron colliders. We hope that this will provide a
new tool for analysis of experimental data from Tevatron experiments
as well as for the development of strategies for the detection of
supersymmetry in experiments at supercolliders.

\vglue 0.4cm
{\elevenbf \noindent\it 1.1 The Minimal Supersymmetric Model Framework}
\vglue 0.4cm

The exploration of strategies by
which sparticles might be detected at high energy hadron colliders, of course,
entails a knowledge of sparticle production cross sections and their decay
patterns. The cross sections for the pair production of sparticles are
essentially determined by their
gauge interactions. Thus, aside from model-dependent mixing angles, these
cross sections are
fixed by their $SU(3)\times SU(2)_L\times U(1)_Y$ quantum numbers.
As a result, the
production cross sections for gluinos, sleptons and squarks
are independent of the details of any
model. This is not the case for the production of charginos and
neutralinos which are model-dependent mixtures of gauginos and Higgsinos.
Sparticle decay modes depend on masses, mixings and kinematically allowed
modes, and so are generally also model dependent.

Here, we adopt the
MSSM as a guide to sparticle masses and mixing angles\cite{MSSM}.
The MSSM is the
simplest supersymmetric extension of the SM. It contains the minimal number of
new particles and interactions consistent with phenomenology. Corresponding
to each chiral matter fermion multiplet there is a spin zero sfermion
multiplet with the same internal quantum numbers.  The superpartners of
the Yang-Mills gauge bosons are the spin $1\over 2$ Majorana gauginos in the
adjoint representation of the gauge group. Finally, any SUSY model requires
at least two Higgs boson doublets to cancel anomalies and to
give mass to both $T_3 = {1\over 2}$
and $T_3 = -{1\over 2}$
fermions: the SUSY partners of these are two spin ${1\over 2}$ doublets
of Higgsinos.
The gauge interactions of the model automatically conserve a discrete
quantum number, the R-parity, which is $+1$ ($-1$) for ordinary
(supersymmetric)
particles. We assume that this is a symmetry of the complete Lagrangian
so that R-parity violating Yukawa type interactions (which necessarily
violate baryon or lepton number) are absent. Optimally, one would like
to include all possible interactions (including R-parity violating ones)
consistent with gauge symmetry. This is phenomenologically unacceptable
since it leads to catastrophic proton decay which can only be avoided
by assuming that at least one of baryon number, lepton number or R-parity
is conserved. We assume the discrete symmetry is R-parity. The most
important consequence of this is that sparticles can only decay into
other sparticles until the decay cascade terminates in the lightest
supersymmetric particle (LSP) which is stable. Because of cosmological
considerations, the LSP is
expected to be a weakly interacting neutral. Hence it is expected to
escape detection in the experimental apparatus, leading to the classic
$\eslt$ signature for supersymmetry. In ISAJET~7.0, the lightest
neutralino $\tz_1$ is assumed to be the LSP; the other possible LSP
candidate, the sneutrino $\tnu$, is heavily disfavored by a combination
of constraints from LEP and dark matter searches\cite{BDT} if we further assume
that the LSP forms the galactic dark matter.

The supersymmetric particles, {\it i.e.} the mass eigenstates of the MSSM,
include the gluinos $\tg$, which being color octet fermions
cannot mix with anything
since $SU(3)_C$ is unbroken, and the spin-zero sfermion partners $\tilde f_L$
and $\tilde f_R$
of the left- and right-handed fermions, whose mixing is proportional to
the corresponding fermion mass and hence is negligible for all but the
top squarks.
Finally, the gauginos and Higgsinos with the same electric
charge mix once $SU(2)_L\times U(1)_Y$ is broken to form two Dirac charginos
$\tw_1$ and $\tw_2$ ($m_{\tw_1} < m_{\tw_2})$
and four Majorana neutralinos $\tz_1 ,\tz_2 ,\tz_3 ,\tz_4$ (in
order of increasing mass). In addition, there are five physical spin
zero bosons associated with the Higgs sector: these are the light and
heavy neutral scalars $H_l$ and $H_h$, a pseudoscalar, $H_p$ (the terms scalar
and pseudoscalar refer to their couplings to matter fermions) and a
pair of charged scalars $H^\pm$.

Most of the sparticle interactions relevant to collider phenomenology
are fixed by the corresponding interaction of their SM partners.
Only one new parameter ($\mu$)\cite{FOOTNOTE},
which corresponds to a supersymmetric mixing
between the two Higgs doublet fields can be introduced. The remaining
supersymmetric interactions can be written in terms of the coupling
constants of the
gauge and Yukawa interactions that are already present in the SM.
Supersymmetry
cannot be exact if it has to be phenomenologically relevant.
Further, if SUSY is to be the resolution of the gauge
hierarchy problem, supersymmetry breaking effects can all be
parametrized by soft SUSY breaking interactions (these are interactions
whose introduction does not lead to the reappearance of quadratic divergences)
all of which have been classified in Ref.~\cite{SOFT}. For our purposes, it is
sufficient to know that both scalar and gaugino mass terms are soft. We
can thus parametrize the breaking of supersymmetry in terms of these SUSY
breaking masses which break the degeneracy between the fermions and their
supersymmetric partners. There are other SUSY breaking interactions
such as the trilinear scalar interaction which is responsible for
$\tilde t_L -\tilde t_R$ mixing; these will be incorporated into
later versions of the program.
It is important to stress that SUSY breaking does not alter the
gauge interactions of the sparticles so that there are observable consequences
of the underlying (softly broken) supersymmetry.

As discussed in Ref.~\cite{ERICE}, the SUSY breaking
masses for each SM multiplet can be independent.
This results in a proliferation of free
parameters that make any phenomenological analyses intractable.
Motivated by supergravity models\cite{SUGRA} in which supersymmetry
breaking effects
in a hidden sector are communicated to the observable sector by universal
gravitational interactions,
we assume here that all the matter sfermions
have a
common mass at the unification scale. Thus, supersymmetry
fixes the Lagrangian at the unification scale
in terms of just a few parameters. In order for us to use this Lagrangian for
perturbative calculations at the 100 GeV scale relevant to experiment today,
these have to be evolved down to the low energy scale using
the renormalization
group\cite{RGE}. The $SU(2)_L$ and
$U(1)_Y$ gaugino masses are then fixed by the gluino mass by the well known
unification condition\cite{MSSM}.
The renormalization group evolution also splits the
degeneracy between the various sfermions. The biggest effect is due to color
interactions so that the largest splitting occurs between squarks and sleptons,
with smaller splittings between the doublet and singlet sfermions. In our
analysis, we have ignored mass splittings between the various squarks.
This is a good approximation except for the third generation sfermions
where the corresponding Yukawa interactions (which, for instance, cause
$\tilde t_L - \tilde t_R$
mixing) can be important. It is primarily
for this reason that $\tilde t$ squarks have
not yet been incorporated into the program.
The slepton masses are also determined
by the common scalar mass, and so are fixed in terms of $m_{\tilde q}$.
Since light sleptons can have a significant impact on neutralino decay
patterns, the D-terms responsible for mass splittings can play an important
role. Although ISAJET / ISASUSY allows
one to input $m_{\tl_L}$, $m_{\tl_R}$
and $m_{\tnu}$ independently, it is quite straightforward to fix these
as in the MSSM framework. Finally, the Higgs
sector of the MSSM is strongly constrained so that it can be specified by just
one additional parameter (which we take to be the mass of $H_p$) which we will
assume is independent of the sfermion mass.

To recapitulate, we see that with the assumptions outlined above,
the masses and couplings of all the sparticles are determined in terms
of just a few parameters which may be taken to be, ({\it i})
the gluino mass which is assumed to determine
the other gaugino masses; ({\it ii}) the squark mass and slepton masses;
({\it iii}) the supersymmetric
Higgsino mass ($\mu$); ({\it iv}) $\tan\beta = v / v'$, the
ratio of the vacuum expectation values of the Higgs fields that couple
to $T_3 = {1\over 2}$ and $T_3 = -{1\over 2}$ fermions.
In addition, $m_{H_p}$ fixes the
tree level masses and couplings of the five Higgs bosons of the MSSM.
Radiative corrections\cite{HIGGS} due to
top quark Yukawa interactions which substantially alter their masses and mixing
patterns (so that these acquire a dependence on the top quark mass as well
as on the other SUSY parameters) have also been incorporated.

The MSSM thus provides a framework for
studying SUSY signals in experiments at current and future colliders. It
should be regarded as an effective low energy theory obtained under certain
reasonable assumptions, which may in the future be derived from a deeper
underlying theory. It should be kept in mind that the six parameters
introduced above may, in fact, be related as, for instance, in
supergravity models with
radiative electroweak symmetry breaking\cite{GRZ}.
The resulting sparticle spectrum can be directly
incorporated into the program. Predictions of No Scale models and
flipped models
can also be obtained as special cases of the general parametrization
of our program. It should be kept in mind that
the predictions of the MSSM should primarily be used
to guide our thinking about strategies for the detection
of supersymmetric particles. While we expect that the qualitative features
of the MSSM predictions are almost certain to be reliable (unless R-parity
is violated), it should be kept in mind that the actual signals may
differ in detail even if nature indeed proves to be supersymmetric. It is
worth remarking that R-parity violating signals from an unstable LSP
can also be studied using this
code by forcing the decay of the LSP as discussed in Sec. 5.

\vglue 0.6cm
{\elevenbf\noindent 2. Physics Content of ISAJET / ISASUSY}
\vglue 0.4cm
{\elevenbf \noindent\it 2.1 Event simulation with ISAJET 7.0}
\vglue 0.4cm

ISAJET\cite{ISA} is a Monte Carlo program which simulates $pp$, $p\bar p$
and to a lesser extent $e^+e^-$ interactions at high energy.
Much of the simulation is based upon perturbative leading-log QCD,
along with phenomenological models for non-perturbative aspects such
as hadronization and beam jet evolution. Event
simulation is carried out by the following steps:
\begin{itemize}
\item calculation of hard scattering subprocess Feynman diagrams,

\item convolution with $Q^2$ dependent parton distribution functions,

\item implementation of approximate all orders QCD corrections via
final and initial state parton showers\cite{FWSG},

\item implementation of predicted particle and sparticle decays,
along with parton radiation and independent
quark and gluon hadronization\cite{FF},

\item suitable modelling of the underlying event structure and beam jet
evolution\cite{AGK}.
\end{itemize}

\noindent More detailed aspects of the above steps are described in
Ref.~\cite{PP}.

To incorporate supersymmetric processes into
ISAJET, the appropriate sparticle subprocess production cross-sections
and the corresponding sparticle decay modes as predicted within the
MSSM framework are needed.
Both production and decay processes depend in general on
the parameter set $m_{\tg}$, $m_{\tq}$, $\mu$, $\tan\beta$,
$m_{H_p}$, and $m_t$. Other elements of the simulation are
essentially unchanged.
The
complete spectrum of MSSM sparticle states have been defined within ISAJET,
with accompanying identification codes.
The supersymmetric particle IDENT codes distinguish between the
partners of left and right handed fermions and include the Higgs
sector of the minimal supersymmetric model:
\begin{verbatim}
   UPSSL, DNSSL, STSSL, CHSSL, BTSSL, TPSSL = 21,22,23,24,25,26
   NUEL, EL-, NUML, MUL-, NUTL, TAUL- = 31,32,33,34,35,36
   UPSSR, DNSSR, STSSR, CHSSR, BTSSR, TPSSR = 41,42,43,44,45,46
   NUER, ER-, NUMR, MUR-, NUTR, TAUR- = 51,52,53,54,55,56
   GLSS = 29
   Z1SS, Z2SS, Z3SS, Z4SS = 30,40,50,60
   W1SS+, W2SS+ = 39,49
   HL0 = 81              HH0 = 82
   HA0 = 83              H+ = 85  .
\end{verbatim}
Anti-particle states of the above are referred to by negative
IDENT codes. The
right sneutrino states of course do not enter the MSSM, and the
left- and right- stop states may be changed in the future to
light and heavy stop mass eigenstates. The \verb|HA0| state refers
to the pseudoscalar Higgs boson; other particle labels ought to be
self-evident.

\vglue 0.4cm
{\elevenbf \noindent\it 2.2 Sparticle production processes}
\vglue 0.4cm

The ${\cal O}(\alpha_s^2 )$ hard scattering subprocesses included
in ISAJET 7.0 are,
\begin{eqnarray}
gg &\to& \tg \tg , \\
q\bar q &\to& \tg \tg , \\
gq &\to& \tg \tq_{i} ,\\
gg &\to& \tq_{i} \tq_{j}, \\
q\bar q &\to& \tq_{i}\bar{\tq_{j}} , \\
qq &\to& \tq_{i}\tq_{j}.
\end{eqnarray}

Since the decay patterns of left- and right-squark types are different,
ISAJET keeps track of squark flavour and type (denoted by the subscripts i
and j). The production cross sections may be found, for example, in
Ref.~\cite{DEQ}. All squark types are currently assumed to be
degenerate in
mass. Top squark pair production is not yet included at this time, so
ISAJET~7.0 generates only 5 flavors of squarks, although it requires
input of $\tst_L$ and $\tst_R$ masses for calculation of various loop decays
and radiative corrections.

The ${\cal O}(\alpha\alpha_s )$ subprocesses which generate squarks or
gluinos in association with charginos or neutralinos are also included
in ISAJET 7.0. The cross sections for the associated production
processes are as
given in \cite{ASSPROD}. The subprocesses, which occur via squark exchange, are
\begin{eqnarray}
g q &\to&\tq \tw_{i}  ,\\
g q &\to&\tq \tz_{i}  ,\\
q\bar q &\to& \tg\tw_{i} ,\\
q\bar q &\to& \tg\tz_{i} .
\end{eqnarray}

Finally, pair production of charginos with neutralinos and
chargino pair production are included. These subprocesses currently
contain only $W$ or $\gamma /Z$ $s$-channel graphs, which should be
the most important ones in the mass range relevant to the Tevatron.
The subprocesses are,
\begin{eqnarray}
q\bar q &\to& \tw_{i} \tz_{j} ,\\
q\bar q &\to& \tw_{i} \bar{\tw_{j}} .
\end{eqnarray}
Explicit formulae
can be readily obtained by modifications of the formulae
in Ref.~\cite{BBKMT}. Neutralino pair
production occurs typically at low rates, and will be included in the future.

The desired subprocesses are chosen by using the ISAJET \verb|JETTYPEi|
keyword commands. The default choice is to generate all allowed
reactions in the appropriate proportions. This may not be the best idea,
however, since a large amount of CPU time can be used trying to sort
through the various subprocesses, especially the various squark pair reactions.

\vglue 0.4cm
{\elevenbf \noindent\it 2.3 Sparticle decay modes}
\vglue 0.4cm

The signals for the production of supersymmetric particles obviously
depend on how these decay. As is well known by
now\cite{BEGNT},
heavy sparticles generically decay into lighter sparticles with the decay
cascade terminating in the (stable) LSP. The branching ratios for the
various sparticle decays as given by the MSSM are computed in the set
of routines labelled ISASUSY.

Since all the sfermions (squarks and sleptons)
have non-trivial gauge interactions, they
can decay into all the neutralinos that are kinematically
accessible in the decays,
\begin{eqnarray}
\tilde{f} &\to& f + \tz_{i}.
\end{eqnarray}
The $SU(2)$ doublet sfermions, $\tilde{f}_L$ can also decay via,
\begin{eqnarray}
\tilde{f}_L &\to& f' + \tw_{i}.
\end{eqnarray}
These decays are forbidden for the singlet sfermions in the limit
that their Yukawa interactions are negligible. The squarks can
also decay by strong interactions via $\tq \to q + \tg$.
This decay dominates if it is not kinematically
suppressed. Since the LSP is assumed to be the lightest neutralino,
the two body decay, $\tilde{f} \to f +  \tz_1$ is always accessible.

The gluinos, being electroweak singlets, can only decay via
\begin{eqnarray}
\tg \to \tq + q
\end{eqnarray}
provided this decay is kinematically allowed; otherwise, the gluino
decays via a virtual squark into a 3-body mode:
\begin{eqnarray}
\tg \to q\bar{q} + \tz_{i}
\end{eqnarray}
or,
\begin{eqnarray}
\tg \to q\bar{q}' + \tw_{i}.
\end{eqnarray}
Explicit partial widths are given in Ref.~\cite{BBKT}.
In the computation of the decay widths
for $\tg \to t\bar{t}\tz_{i}$ and $\tg \to t\bar{b}\tw_{i}$, we have used
the formulae in Ref.~\cite{LOOP,MULTILEP} where the effects of the top family
Yukawa interactions are included.  Notice that the decay
$\tg \to q\bar{q} +\tz_{1}$ is always allowed.
Finally, we note that we have also included
the two  body decay,
\begin{eqnarray}
\tg \to g + \tz_{i}
\end{eqnarray}
which can be important\cite{LOOP} in certain regions of the parameter
space.

The charginos and neutralinos, if they are heavy enough, can decay
via two body modes,
\begin{eqnarray}
\tw_{i} &\to& \tz_{j} + (W\ {\rm or}\  H^\pm ) , \\
\tw_2 &\to& \tw_1 + (Z\ {\rm or}\ H_l,\ H_h,\ H_p ) , \\
\tz_{i} &\to& \tz_{j} + (Z\ {\rm or}\ H_l,\ H_h,\ H_p ) ,
\end{eqnarray}
and,
\begin{eqnarray}
\tz_{i} &\to& \tw_{j} + (W\ {\rm or}\ H^\pm ) .
\end{eqnarray}
If sfermions are light enough, the decays
\begin{eqnarray}
\tz_{i} &\to& \tf_{j} + \bar{f}_j
\end{eqnarray}
and
\begin{eqnarray}
\tw_{i} &\to& \tf_{L} + \bar{f}'_L
\end{eqnarray}
may also be kinematically accessible. Here, we have, as before, assumed
that Yukawa interactions are negligible; otherwise chargino decays
to $\tf_{R}$ would also be possible, as would be decays to Higgs bosons.
Charginos and neutralinos will also decay via a variety of three
body modes,
\begin{eqnarray}
\tw_{i} &\to& f\bar{f}'\tz_{j}, \\
\tz_{i} &\to& f\bar{f}\tz_{j}.
\end{eqnarray}
that are mediated by virtual $W$ or $Z$ bosons and virtual sfermions.
It is worth noting that the inclusion of sfermion mediated neutralino
decay amplitudes including mass splittings between squarks and sleptons
can be very important\cite{BT} for neutralino decay patterns because
the $Z\tilde{Z_i}\tilde{Z_j}$ coupling can be dynamically suppressed.
Finally, we have also incorporated
the decays,
\begin{eqnarray}
\tz_{i} &\to& \tw_{1}f\bar{f}'
\end{eqnarray}
into the program.

Although the direct production of the Higgs bosons has not yet been
incorporated into ISAJET, these can be produced via the cascade decays
of sparticles as discussed above. The charged Higgs boson can also be
produced via the decay $t \to bH^+$ whenever it is kinematically
accessible. In order to provide a complete simulation of SUSY events,
the various decay modes of the MSSM Higgs bosons have, therefore, been included
into ISASUSY. In our computation, we have included the effects
of radiative corrections
due to top quark Yukawa couplings\cite{HIGGS} using the formulae in
Ref.~\cite{BBDKT};
we have not yet included radiative corrections
from the bottom quark Yukawa interactions (which only become important for
very large values of $\tan\beta \gg 1$)
or from gauge interactions. We have included the tree level
decays\cite{HHG} of neutral Higgs bosons to SM particles,
\begin{eqnarray}
 H_l,\ H_h\ {\rm or}\ H_p &\to& f\bar{f},
\end{eqnarray}
\begin{eqnarray}
 H_l,\ H_h &\to& VV\ {\rm or}\ VV^* ,\ (V = W, Z)
\end{eqnarray}
as well as the one loop decays,
\begin{eqnarray}
 H_l,\ H_h,\ H_p &\to& \gamma\gamma\ {\rm or}\ gg .
\end{eqnarray}
In addition, we have included decays to chargino and neutralino
pairs\cite{BBKMT},
\begin{eqnarray}
H_l,\ H_h\ {\rm or}\ H_p &\to& \tz_{i}\tz_{j}\ {\rm or}\ \tw_{i}\tw_{i}
\end{eqnarray}
whenever these are kinematically allowed, as well as the decays,
\begin{eqnarray}
H_h &\to& H^+ H^-,\ H_l H_l,\ H_p H_p\ {\rm or}\ H_p Z, \\
H_p &\to& H_l Z, \\
H_l &\to& H_p H_p.
\end{eqnarray}
of MSSM Higgs bosons into lighter Higgs bosons.

For the charged Higgs bosons, we have incorporated the decays,
\begin{eqnarray}
H^\pm \to f\bar{f}',\ W H_l\ {\rm or}\ \tw_{i}\tz_{j}
\end{eqnarray}
into ISASUSY. The decays of both charged and neutral Higgs bosons into
sfermion pairs have yet to be incorporated.

\vglue 0.5cm
{\elevenbf \noindent 3. Setting up and running ISAJET~7.0 / ISASUSY~1.0}
\vglue 0.4cm
{\elevenbf \noindent\it 3.1 ISAJET 7.0}
\vglue 0.4cm

The ISAJET package has been encoded using the PATCHY code
management system developed at CERN. The file ISAJET.CAR containing
source code and PATCHY commands is available by copying directly
from the Brookhaven VAX. For instance, on VMS
\begin{verbatim}
$ copy BNLCL6::$2$DUA14:[ISAJET.ISALIBRARY]ISAJET.CAR *
\end{verbatim}
can be used to obtain the source code. Ample disk space should be procured
beforehand since the code is $\sim 36,000$ lines long.

Simple programs using PATCHY commands can then be used to assemble
the appropriate FORTRAN files, decay table and documentation.
For example, on a VAX, running a .COM file including the following commands
\begin{verbatim}
$ YTOBIN
ISAJET ISAJET - - YTOBIN .GO
$ YPATCHY
ISAJET ISATEXT.TXT TTY YPATCHY .GO
+USE,*ISATEXT,VAX
+EXE
+PAM
+QUIT
$ YPATCHY
ISAJET ISAJET TTY YPATCHY .GO
+USE,*ISAJET,VAX
+USE,*ISAPLT
+USE,IMPNONE,NOCERN
+EXE
+PAM
+QUIT
$ YPATCHY
ISAJET ISADECAY.DAT TTY YPATCHY .GO
+USE,*ISADECAY
+EXE
+PAM
+QUIT
\end{verbatim}
will assemble the appropriate ISAJET.FOR file, the decay table, and
the ISAJET documentation file. The ISAJET files may then be inserted
into a library to be linked with main calling programs.
To assemble ISAJET on other machines, or to assemble patches such as
ISAZEB or ISAPLT, see Ref.~\cite{PP} and the generic UNIX Makefile
MAKEFILE.UNIX available from BNLCL6.

A main program to run ISAJET can be simply constructed:
\begin{verbatim}
      PROGRAM MSSM
      OPEN(UNIT=1,FILE='ISADECAY.DAT',STATUS='OLD',FORM='FORMATTED')
      OPEN(UNIT=2,FILE='ISAJET.DAT',STATUS='NEW',FORM='UNFORMATTED')
      OPEN(UNIT=3,FILE='ISAJET.PAR',STATUS='OLD',FORM='FORMATTED')
      CALL ISAJET(-1,2,3,6)
      STOP
      END
\end{verbatim}
After compiling, one may link with the ISAJET library of routines,
using for VMS
\begin{verbatim}
$ LINK MSSM,ISAJET/LIB/INCL=ALDATA
\end{verbatim}

To run with the above main program, the decay table ISADECAY.DAT
created above must be included in the
user's directory, as well as a file of commands ISAJET.PAR, which specifies
crucial input commands.

An example of an ISAJET.PAR file is
\begin{verbatim}
SAMPLE CHARGINO PAIR JOB AT TEVATRON
1800,1000,2,500/
SUPERSYM
BEAMS
'P','AP'/
MSSM1
200,250,250,250,250,250,250/
MSSM2
2,500,-100/
JETTYPE1
'W1SS-','W1SS+'/
JETTYPE2
'W1SS-','W1SS+'/
PT
20,50,20,50/
END
STOP
\end{verbatim}
In the above file, the first line is simply the program title. The second
line contains the machine energy \verb|(Ecm)|, the
number of events to be generated \verb|(Nevent)|,
the number of events to be output to screen \verb|(Nprint)|, and
how many events to skip before
printing another event to screen \verb|(Njump)|.
The first event is always printed if \verb|Nprint| is greater than zero.
The third line specifies the reaction type,
which for supersymmetry is always \verb|SUPERSYM|, although many other
non-supersymmetric options are available (see Ref.~\cite{PP}.) Next
come double lines containing a {\it keyword} and then {\it input}. First,
we specify beam types proton and anti-proton. We next specify the
keywords \verb|MSSM1| and \verb|MSSM2|.
The input for \verb|MSSM1| is $m_{\tg}$, $m_{\tq}$, $m_{\tst_L}$,
$m_{\tst_R}$, $m_{\tl_L}$, $m_{\tl_R}$, and $m_{\tnu_L}$. The input for
\verb|MSSM2| is $\tan\beta$, $m_{H_p}$ and $\mu$.
All mass dimension parameters are in GeV units. Invoking the \verb|MSSM1|
and \verb|MSSM2| keywords causes the ISASUSY decay package to be called,
so that all sparticle masses, decays modes and branching fractions
are calculated, and entered into ISAJET's internal decay table. Note
{\it both}
\verb|MSSM1| and \verb|MSSM2| must be specified to use ISASUSY.

Next, one must specify the two final state particles of the $2\to 2$ hard
scattering. These are specified by the keywords \verb|JETTYPE1| and
\verb|JETTYPE2|, using \verb|A8| input format in single quotes.
Currently available possibilities include:
\begin{verbatim}
'GLSS',
'UPSSL','UBSSL','DNSSL','DBSSL','STSSL','SBSSL','CHSSL','CBSSL',
'BTSSL','BBSSL','TPSSL','TBSSL',
'UPSSR','UBSSR','DNSSR','DBSSR','STSSR','SBSSR','CHSSR','CBSSR',
'BTSSR','BBSSR','TPSSR','TBSSR',
'W1SS+','W1SS-','W2SS+','W2SS-','Z1SS','Z2SS','Z3SS','Z4SS',
'SQUARKS','GAUGINOS'.
\end{verbatim}
The last two generate respectively all allowed combinations of squarks
and antisquarks and all combinations of charginos and neutralinos, with
proportions as given by the MSSM.
Care must be taken in specifying \verb|JETTYPEi|. For instance, in the
above example, the output total cross-section would correspond to the expected
total cross-section for chargino pair production. If we had instead
specified
\begin{verbatim}
JETTYPE1
'W1SS-'/
JETTYPE2
'W1SS+'/
\end{verbatim}
the final cross-section tally would be only {\it half} the total cross-section,
although all events and distributions would be correctly generated.
The other half of the cross section would be obtained by interchanging
the two \verb|JETTYPEi| arguments.
Finally, in the above example, the $p_T$ limits of the final state
particles of the $2\to 2$
hard scattering subprocess are specified as
($p_{T_1}(min),p_{T_1}(max),p_{T_2}(min),p_{T_2}(max)$).
For more on keyword options, see
Ref.~\cite{PP}.

\vglue 0.4cm
{\elevenbf \noindent\it 3.2 ISASUSY 1.0}
\vglue 0.4cm
ISASUSY 1.0 --- now a subset of ISAJET 7.0 --- is used to
calculate sparticle masses, mixings and branching fractions. ISASUSY~1.0
is automatically called by ISAJET~7.0 whenever the \verb|MSSM1| and
\verb|MSSM2| keywords are used. In this case, ISASUSY fills an internal
ISAJET decay table with the appropriate decay modes and branching fractions;
the modes themselves are not printed since output consists of many pages.
The user may however run ISASUSY 1.0 as a separate package to generate
a file of all calculated masses and decay modes, partial widths and
branching fractions.

To assemble ISASUSY.FOR for independent runs,
the following VMS command file can be used:
\begin{verbatim}
$ YTOBIN
ISAJET ISAJET - - YTOBIN .GO
$ YPATCHY
ISAJET ISASUSY TTY YPATCHY .GO
+USE,*ISASUSY
+USE,VAX,IMPNONE,NOCERN
+EXE
+PAM
+QUIT
\end{verbatim}
Compiling and linking are straightforward, since ISASUSY doesn't need to
be linked with any other files.

When running ISASUSY, the program will ask for an output filename in
single quotes. After entering, ISASUSY asks for the same parameter set
as ISAJET: $m_{\tg}$, $m_{\tq}$, $m_{\tst_L}$, $m_{\tst_R}$,
$m_{\tl_L}$, $m_{\tl_R}$, $m_{\tnu_L}$, $\tan\beta$, $m_{H_p}$, $\mu
$, $m_t$.
Output will then be written to the specified file for viewing or
printing.

It should be noted that some choices of parameters will result in
$m_{\tw_1}<m_{\tz_1}$, violating the assumption that $\tz_1$ is the
LSP. In this case, ISASUSY replies with a warning, and terminates
execution. Other choices of parameters can be in regions already
excluded by LEP constraints. At present, no warnings are issued for
this case.

\vglue 0.5cm
{\elevenbf \noindent 4. ISAJET Output}
\vglue 0.4cm
Upon generating events, ISAJET fills various COMMON blocks listed in
Ref.~\cite{PP}. Explicit output of all beginning run information, all the
events, and end run information is stored in the file ISAJET.DAT in the
sample program given in Sec.~3. To access this information, one can read
the file ISAJET.DAT using the RDTAPE subroutine of ISAJET. For instance,
the following program
\begin{verbatim}
      PROGRAM READ
      COMMON /RECTP/ IRECTP,IREC
      COMMON/ITAPES/ ITDKY,ITEVT,ITCOM,ITLIS
      ITLIS=6
      OPEN(UNIT=1,NAME='ISAJET.DAT',TYPE='OLD',FORM='UNFORMATTED')
10    CALL RDTAPE(1,IFL)
      IF (IFL.NE.0) GO TO 20
      IF (IRECTP.EQ.100) THEN
        CALL PRTEVT(0)
      END IF
      GO TO 10
20    STOP
      END
\end{verbatim}
will open and read the file ISAJET.DAT, and restore event information
to the ISAJET common blocks so that it can be manipulated, if the user
inserts the appropriate common blocks. This program
also prints the event information to the terminal screen. Information
on the final total cross section can be found in common block
\verb|/FINAL/|.
\vglue 0.5cm
{\elevenbf \noindent 5. Extensions and Future Improvements}
\vglue 0.4cm
ISAJET 7.0 contains sufficient flexibility that some scenarios for
non-minimal SUSY can also be studied. For instance, R-violating
models with an unstable $\tz_1$ can be easily simulated by using the
ISAJET \verb|FORCE| keyword command to force the desired $\tz_1$ decay.
\verb|FORCE| can also be used to override ISASUSY generated decays,
or to select specific decay modes for certain sparticles.
In addition, the unification condition on gaugino masses can be relaxed
if desired. This can be done by a simple modification of the
FORTRAN code in subroutine SSMASS. Some modifications, such as
added Higgs singlets which can also enlarge the neutralino sector, are
more difficult to include, and would require a more substantial code revision.

ISAJET 7.0 can also be used to simulate top squark events
even before stops are officially included. For instance, by
setting the \verb|JETTYPEi| to be bottom squarks, one generates nearly the
same cross-section as for top squarks. Then the user may use the
\verb|FORCE| command to force the generated squarks to decay into the
desired modes, such as $\tst_1\to c\tz_1$ or $\tst_1\to b\tw_1$
(see Ref.~\cite{BDGGT}).

There are still many aspects of MSSM sparticle production that are
not included in the current version of ISAJET, but will hopefully be
included in future versions. A partial list includes the following:
\begin{itemize}
\item top squark production and decay;
\item slepton pair production processes;
\item a subroutine to notify if the parameters are in violation of LEP
limits;
\item direct MSSM Higgs boson production mechanisms;
\item neutralino pair production processes;
\item Higgs decays to sfermion pairs;
\item further breaking of sfermion degeneracies, especially for the
third generation;
\item improved calculations of radiative corrections to the Higgs
sector;
\item $e^+ e^-$ production of SUSY particles.
\end{itemize}

\vglue 0.5cm
{\elevenbf\noindent 6. Acknowledgements\hfil}
\vglue 0.4cm

We thank M. Bisset and D. Dzialo (Karatas) for
programming input and collaborations. We also thank
J. Freeman, S. Hellman, C. Kao,
M. Paterno, F. Pauss, G. Polesello, B. van Eijk, A. White,
J. White and J. Woodside for their contributions by way of collaboration,
encouragement, and detection of many bugs in early
stages of this work. We are also grateful to many colleagues for
collaborations and discussions through the years.
This research was supported in part by the US Dept. of Energy Contract No.
DE-FG05-87ER40319 and DE-AM03-76SF00235 and by the TNRLC SSC Fellowship
program.

\vglue 0.5cm
{\elevenbf\noindent 7. References \hfil}
\vglue 0.4cm

\end{document}